\documentstyle[eqsecnum,epsf,aps,prb,multicol]{revtex}
\begin{document}
\draft

\title{Dual Order Parameter for the Nodal Liquid}
\author{Leon Balents}
\address{Bell Labs, Lucent Technologies, 700 Mountain Ave, Murray
  Hill, NJ 07974}
\author{Matthew P. A. Fisher}
\address{Institute for Theoretical Physics, University of
California, Santa Barbara, CA 93106-4030}
\author{Chetan Nayak}
\address{Physics Department, University of California, Los Angeles, CA 
  90095--1547}
\date{\today}
\maketitle

\begin{abstract}
The guiding conception of vortex-condensation-driven
Mott insulating behavior is
central to the theory of the nodal liquid.
We amplify our earlier description of this idea
and show how vortex condensation in $2D$ electronic
systems is a natural extension of $1D$ Mott insulating
and $2D$ bosonic Mott insulating behavior.
For vortices in an underlying superconducting pair field,
there is an important distinction between
the condensation of flux $hc/2e$ and flux $hc/e$ vortices.
The former case leads to spin-charge confinement,
exemplified by the band insulator and the charge-density-wave.
In the latter case, spin and charge are liberated
leading directly to a $2D$ Mott insulator
exhibiting {\it spin-charge} separation.
Possible upshots include not only the nodal liquid, but also
a novel undoped antiferromagnetic insulating phase
with gapped excitations exhibiting spin-charge
separation.

\end{abstract}
\vskip 0.5 in
\begin{multicols}{2}

\section{Introduction}

The present paper is rooted in the conviction that the basic property
of an insulator is that it insulate.  Magnetic order is a secondary
effect which, though it may be one of the appurtenances of insulating
behavior, is not synonymous with it. This premise underlies our recent
discussion of the phase diagram of the underdoped cuprate
superconductors, where we relied upon the notion of an insulator as a
vortex condensate\cite{BFN}.  We would like, here, to elucidate and expand upon
this paradigm. As a by-product of this approach, we find a precise
distinction between spin-charge separated and confined insulators.

Since correlated insulators often exhibit magnetism, they are
typically described by their magnetic order parameters.  Furthermore,
in commensurate, weak coupling models, the development of magnetic
order can be the mechanism by which a system becomes insulating. This
has led to the conflation of magnetic order with insulating behavior.
However, this state of affairs is both unsatisfying and incomplete
since magnetic order can persist even when the system becomes
conducting and, conversely, a system can be insulating even in the
absence of magnetic order, as exemplified by the nodal liquid. Hence,
it would behoove us to seek an order parameter which is directly
related to the electrical properties of an insulator.  The difficulty
in such a program is that an insulator seems more disordered than a
conductor since most correlation functions in the charge sector are
short-ranged.  We interpret this as suggesting a `dual' approach based
on a `disorder' parameter.  Distilling the key elements of our nodal
liquid construction, we propose that the appearance of a non-zero
expectation value for a `disorder' parameter signals insulating
behavior.

The relation between the conduction properties of a system and
spontaneous symmetry breaking in a dual order parameter is perhaps
most transparent in the field-theoretical formulation of
duality.\cite{ftduality}\ In this version of the transformation, the
dual theory is constructed to implement local charge continuity, which
is a {\sl dynamical} consequence of U(1) symmetry in the original
Hamiltonian, as a rigid {\sl constraint} (dynamics in the dual theory
implies conservation of {\sl vorticity}).  This construction is quite
familiar from the one-dimensional theory of Luttinger liquids\cite{Lutt}.
There,
the two-current can be written in terms of a phase field
${\phi}(x)$ as
\begin{equation}
  {j_i} = {\partial_i}{\phi} .
\end{equation}
In the alternative, but equivalent, dual description, the current can
be written in terms of a dual field, ${\theta}$:
\begin{equation}
  {j_i} = {\epsilon_{ij}}{\partial_j}{\theta} .
  \label{thetafield}
\end{equation}
In a one-dimensional Mott insulator, the $\theta$ field acquires
a mass, implying a gap in the spectrum of charge excitations, and hence,
experimentally, in the optical conductivity.  Note that massive
dynamics for the {\sl phase} field $\phi$ is inconsistent since
it would violate charge conservation $\partial_i j_i = 0$.  The
one-dimensional model has thus communicated an important lesson:
insulating behavior occurs when a gap is acquired by the dual field
representing the current operator.

This description of Mott insulators generalizes readily to
two-dimensional {\sl bosons}.  Let us write the
boson annihilation operator, $\psi$,
in terms of its amplitude and phase:
$\psi = \sqrt{\rho}{e^{i\varphi}}$.
In the superfluid state, we fix
$\rho$ and describe the system by
its phase degree of freedom, $\varphi(x)$.
\begin{equation}
  {\cal L} = \frac{1}{2}\,\left({\partial_\mu}\varphi\right)^2.
\end{equation}
We can model the destruction of superfluidity by introducing a vortex
field, ${\Phi}$; effectively, we are reducing the amplitude degree
of freedom to a vortex field which keeps track of the points at which
it vanishes.  We can represent the current, ${j_\mu} =
{\partial_\mu}\varphi$, as\cite{ftduality,Duality}
\begin{equation}
  {j_\mu} = {\epsilon_{\mu\nu\lambda}}
  {\partial_\nu}{a_\lambda},
  \label{dualcurrent}
\end{equation}
which is the natural extension of Eq.~\ref{thetafield}
to two-dimensions.
This parameterization is highly redundant as a result of
its invariance under the gauge symmetry:
\begin{equation}
{a_\mu}\rightarrow{a_\mu}+{\partial_\mu}\chi,
\label{gauge}
\end{equation}
for an arbitrary function, $\chi$.  This gauge symmetry is
enormously larger than the analogous global  invariance in the
one-dimensional Luttinger liquid, $\theta \rightarrow \theta
+ {\rm const}$.   The vortices see the
gauge field, $a_\mu$, according to the Magnus force law, so
the dual Lagrangian takes the form
\begin{equation}
  {\cal L}_D = \frac{1}{2}\,{\left({\partial_\mu}{a_\nu}\right)^2}
  + {\left| \left({\partial_\mu}-i{a_\mu} \right)
      {\Phi}\right|^2} + V({\Phi}).
  \label{dualbosons}
\end{equation}
In the superfluid state, vortices are gapped and
the bosons condense, while in the Mott insulating
state, the vortices condense and the bosons become
gapped (at the chemical potential).
When the vortices condense, $a_\mu$
becomes massive by the Anderson-Higgs mechanism.
As a consequence, the system is insulating.

Let us abstract away the archetypal features of this system.  We can
introduce the representation (\ref{dualcurrent}) for any conserved
current in two-dimensions, so we can certainly use it for the 
electrical current in the
fermionic system of our choice.  In order to carry the rest of this
scheme over to electronic systems, we must find a way for the system
to conspire to make $a_\mu$ massive. The only way that this can happen
which is consistent with the gauge symmetry (\ref{gauge}), is via the
Anderson-Higgs mechanism.\cite{foot1}\ Since the Anderson-Higgs
phenomenon can only take place if there is a condensate which is
coupled to $a_\mu$ according to the Magnus force law
(Eq.~ \ref{dualbosons}), we are led to the following question: how do
you
define a vortex field in a fermionic system?  One possibility is to
implement statistical transmutation to represent the fermions as
bosons coupled to an auxiliary Chern-Simons gauge field\cite{Lutt}. Then, we can
define vortices in the bosonic field.
This approach is probably suitable for
describing a conventional antiferromagnet,
as discussed
very briefly in Section IV.
But in this paper we pursue
a different tack - using Cooper pairs as the bosons.
This is quite promising for the cuprates because it is tailor-made for
insulators which contain the germ of superconductivity.

A new question rears its head when we consider an insulator which
descends in this way from a superconductor: do the finite-energy
excitations inherit their quantum numbers from the superconducting
state, or do they simply have the electron quantum numbers? In
particular, one can ask what is the energy of an isolated neutral
spin-$1/2$ excitation. If this diverges with system size, then spin
and charge are confined. If, on the other hand, it is finite, as it is
in a superconductor, then the insulator exhibits spin-charge
separation.  A band insulator is, of course, a state of the former
variety. As we show in the next section, it can be understood
(rather differently than in elementary
textbooks)
as resulting from the condensation of flux $hc/2e$ vortices in a state
with $s$-wave pairing. Spin and charge are confined as a result of the
Aharonov-Bohm phase which a spinful excitation acquires as it orbits an
$hc/2e$ vortex; an isolated spin-$1/2$ has logarithmically divergent
energy in $2D$. This spin-charge confinement physics is also present in
a
charge-density-wave (CDW) insulator, which occurs
for example in the negative $u$ (extended) Hubbard model at
half-filling.
There, however, the
$hc/2e$ vortex condensation leads to translational symmetry-breaking.
In both instances the resulting insulating phase
can be viewed as a ``crystal" of charge $2e$
``Cooper" pairs.  For $2D$ electronic systems at or near half-filling
with strong
on-site repulsion, however, such CDW order is physically
unreasonable, and for the cuprate materials can
be discarded on phenomenological grounds.
For this reason we are led to consider
the possibility of the condensation of
double strength $hc/e$ vortices in a superconducting
pair field\cite{sachdev}.

The Mott insulator which arises at half-filling
upon condensation of $hc/e$ vortices has a number
of appealing and remarkable properties.
The insulating phase is translationally
invariant, in contrast to the CDW.
Moreover, since
there are no Aharonov-Bohm
phase factors when a spin $1/2$ is transported
around an $hc/e$ vortex, the resulting Mott
insulator exhibits {\it spin-charge} separation\cite{Anderson,krs,Affleck}.
For a d-wave superconductor appropriate
to the cuprates, condensation of $hc/e$ vortices
leads directly to the {\it nodal liquid}.
The nodal liquid
indeed posseses {\it gapless}
spin $1/2$ but charge neutral fermions - the nodons - which descend
directly from the d-wave quasiparticles.  As we shall see,
there are also
massive charge $e$ spinless
boson excitations in the nodal liquid phase.  These ``holons"
are solitonic {\it topological} excitations
in the underlying $hc/e$ condensate, a dual analog
of Abrikosov flux tubes.  The excitations in the nodal liquid
have the same quantum numbers as in the spin-charge separated
gauge theories\cite{PALee}, but are weakly interacting,
rather than strongly coupled by a gauge field.

A peculiar feature of the nodal liquid is that
spin-charge separation survives the ordering of
the nodal spins into a phase with long-ranged
antiferromagnetic order.  This novel phase - denoted AF* -
which has gapped nodons
is distinct from the conventional
N\'eel antiferromagnet, AF, which does not have neutral,
spin $1/2$ excitations even at high energies\cite{gedanken}. These two phases are
physically
very different, as may be seen from simple
{\it gedanken} experiments which make the point
that charge $e$ can be physically separated
from spin $1/2$ with finite energy cost
in the AF* but not the AF phase.
However, in two-dimensions nodon-holon bound states
form in the AF* phase, so the spin-charge separation
is not so easily found in the electron spectral function.

Our ultimate goal is to describe a novel spin-charge
separated state, the nodal liquid, and an ordered
state which can result from it, AF*.  Along the way, however,
we will re-examine a number of seemingly quotidian states such as the
band insulator, the charge-density-wave, and the antiferromagnet.
Our broader framework will enable us to
understand the physics of doping these insulators from the point of
view of creating topological excitations in the `disorder' parameter
\cite{tdoping}.
Such a point of view naturally leads to a discussion of the possibility
of spin-charge separation in these states.   In sections II
and IIIA, we will illustrate the physics of
flux $hc/2e$ condensation in systems
with attractive electron-electron interactions, where we expect the
insulating
states to be related to $s$-wave superconductivity. In the
resulting states, the band insulator and the CDW, spin and charge are
confined, as we discuss at length in section II.
In section IIIB, we then consider flux $hc/e$ vortex condensates in a
$d$-wave superconductor, filling in a gap in our earlier paper.  We
introduce an effective lattice model in section IIIC
which incorporates this physics. In section IV, we
discuss the spin-charge separated antiferromagnet, AF*,
and compare it to the conventional antiferromagnet, AF.
Possible experimental signatures are analyzed.  We conclude with some
summary remarks in section V,  relegating some supporting technical
details to the appendix.

\section{The Band Insulator and Spin-Charge Confinement}

In the absence of electron interactions,
a band insulator with two electrons per unit cell
corresponds to a filled valence band of non-interacting
levels.  Provided the interactions are small
compared to the energy gap to the conduction band
this should provide a good description of the phase.
But even with stronger interactions a band insulator
can be adiabatically deformed (without gap closure)
back to the non-interacting state.
To obtain an order parameter for the band insulator,
we will attempt to describe this phase as
a ``quantum disordered" s-wave superconductor.

To this end, consider spinful electrons moving in the two-dimensional
continuum.  In the presence of a local attractive interaction
the Fermi surface is unstable, and presumed to form a spin-singlet
s-wave superconducting phase, with energy gap $\Delta$.
Now imagine introducing a periodic potential with magnitude $V$
and a period corresponding to two electrons per unit cell.
For $V$ much smaller than $\Delta$ the superconducting phase
should be stable, but with increasing $V$ one expects
an eventual quantum transition into a band insulator.  This
transition can be described as a
quantum vortex unbinding transition, analogous
to the thermally driven Kosterlitz-Thouless transition.
At zero temperature the vortices can condense, giving one
an order parameter for the band insulator.

Of particular
interest is the nature of the gapped excitations in the two phases.
As we shall discuss, in a superconductor
it is possible to define charge neutral
quasiparticles\cite{Kivelson90}, which carry spin $1/2$.
In an s-wave superconductor these ``spinon"
excitations are fully gapped\cite{krs}, but can be gapless
(called nodons) in a d-wave superconductor which we return to in Section
IV below.
Since the Cooper pairs carry no spin in a singlet superconductor,
the Cooper pairs and spinons provide a natural spin-charge
separated description of the superconducting phase.
On the other hand, the excitations in a band insulator
are electrons which of course carry both spin and charge - the
band insulator does {\it not} exhibit spin-charge separation.
We would like to try and understand the mechanism whereby
the separated spin and charge excitations in the superconductor
become ``confined"
upon entering the band insulator.

To address these issues
it is convenient to consider a low energy effective
theory for an s-wave superconductor in which
both Cooper pairs and the gapped quasiparticle
states near the
Fermi surface are retained.  The appropriate Lagrangian takes the form:
${\cal L} = {\cal L}_c + {\cal L}_\varphi + {\cal L}_{int}$
with
\begin{equation}
{\cal L}_c = c^\dagger_\alpha(x,\tau) [ i \partial_t - \nabla^2
 - \mu ] c(x,\tau)   ,
\end{equation}
\begin{equation}
{\cal L}_\varphi = {\kappa_\mu \over 2} (\partial_\mu \varphi)^2 ,
\label{lphi}
\end{equation}
\begin{equation}
\label{anom}
{\cal L}_{int} = |\Delta | e^{i\varphi} c_\uparrow(x) c_\downarrow(x) +
h.c.  .
\end{equation}
Here $c_\alpha$ denotes an electron with spin $\alpha$,
and $\varphi$ is the phase of the pair field, with magnitude
$|\Delta|$.  Integrating over high energy electron states
well away from the Fermi surface, will generate
dynamics for the phase field.   The appropriate
form of ${\cal L}_\varphi$ at low energy is essentially determined by
symmetry.
Here we have retained the leading order terms in a gradient expansion,
with $\kappa_0 = \kappa$ the compressibility and $-\kappa_j=v_c^2
\kappa_0$
a superfluid stiffness.  Henceforth we will set the velocity $v_c=1$.
In general a Berry's phase term\cite{Lutt} of the form
${\cal L}_{\rm Berry} =  n_0 \partial_t \varphi$ is allowed (see section III
below) but with one Cooper pair (two electrons) per unit
cell of the periodic potential $n_0 = 1$ and the Berry's phase
term can be dropped since
$\exp(i \int dt {\cal L}_{\rm Berry})=1$.

Notice that the phase fluctuations
are strongly coupled to the electron operators through ${\cal L}_{int}$.
To decouple this interaction and to exhibit the spin-charge separation
it is convenient to consider the change of variables:
\begin{equation}
f_\alpha(x) = e^{i\varphi/2} c_\alpha(x)   .
\label{nodondef}
\end{equation}
The Lagrangian becomes,
\begin{equation}
{\cal L} = {\cal L}_f + {\cal L}_\varphi + {1 \over 2}
J_\mu \partial_\mu \varphi  ,
\end{equation}
with \begin{equation}
{\cal L}_f =
f_\alpha^\dagger [i\partial_t - \nabla^2
 - \mu ] f_\alpha +  |\Delta | f_\uparrow(x) f_\downarrow(x) + h.c.  ,
\end{equation}
and $J_\mu$ is a quasiparticle 3-current operator:
\begin{equation}
J_0 =  f^\dagger_\alpha f_\alpha   ;  \hskip0.5cm
\vec{J} =  i f^\dagger_\alpha \vec{\nabla}
f_\alpha + h.c. .
\end{equation}
This current is not conserved as a result of the anomalous
term (\ref{anom}), but the spin currents,
\begin{equation}
{J_0^i} =  f^\dagger_\alpha \sigma^i_{\alpha\beta} f_\beta   ;
\hskip0.5cm
\vec{J}^i  =  i f^\dagger_\alpha \sigma^i_{\alpha\beta}\vec{\nabla}
f_\beta + h.c. .
\end{equation}
are conserved.
Here we have assumed that the phase field is slowly varying and have
dropped
terms involving two spatial gradients of $\varphi$.

The Lagrangian ${\cal L}_f$ can be diagonalized
as usual by a Bogoliubov transformation, and describes
gapped quasiparticles.  Since the $f$ operators are electrically
neutral but carry spin one-half, these
excitations are ``spinons".
The spinons are coupled to the phase fluctuations via
a Doppler shift type term.  In the superconducting phase
these phase fluctuations are small, and will
generate a weak interaction between the gapped spinon states.

To quantum disorder the superconducting phase and arrive at
a description of the band insulator, we will need to allow for vortices
in the phase of the pair field.
In two-dimensions vortices are
simply whorls of current
swirling around a core region.
The circulation
of such vortices is {\it quantized},
since upon encircling the core
the phase $\varphi$ can only change
by integer multiples of $2\pi$.  The ``elementary"
vortices have a phase winding of $\pm 2\pi$.
Since the Cooper pairs have charge $2e$,
in the presence of an applied magnetic field
these vortices quantize flux in units of $hc/2e$.
Inside the core of a vortex the {\it magnitude}
of the complex order parameter $|{\Delta}|$ vanishes,
but is essentially constant outside.
Since the position of these ``point-like" vortices
can change with time, their dynamics
requires a quantum mechanical description.
Thus a collection of many vortices can be viewed as a
many body system of ``point-like" particles.  Since
positive ($+1$) and negative ($-1$) circulation vortices can
annihilate and disappear they behave
as ``relativistic" particles.  There is a
conserved vortex ``charge" in this process, namely the total
circulation, and an associated current.
A duality transformation can be implemented\cite{ftduality,Duality}
in which the phase $\varphi$ is replaced
by a dual field, $\theta$, which is the phase
of a vortex complex field, $\Phi \sim e^{i\theta}$.
In a Hamiltonian description,
$\Phi$ and $\Phi^\dagger$ can be viewed as vortex quantum field
operators - which destroy and create vortices.

A crucial element in the duality transformation is
the total electrical three-current,
$J^{tot}_\mu = \kappa_\mu \partial_\mu \varphi + J_\mu/2$,
which
must be conserved even in the dual representation.
This is achieved by expressing
the current as a curl of a gauge field, $a_\mu$\cite{ftduality}:
\begin{equation}
J^{tot}_\mu = \epsilon_{\mu\nu\lambda} \partial_\nu a_\lambda  ,
\end{equation}
which automatically implies the continuity
equation $\partial_\mu J^{tot}_\mu =0$.
This representation also introduces a gauge symmetry
into the problem, $a_\mu \rightarrow a_\mu + \partial_\mu \Lambda$.
It is this gauge symmetry which is spontaneously
broken in the band insulating state.

In Ref. \onlinecite{BFN} the duality transformation was implemented
in the presence of the Doppler-shift interaction between
the Cooper pairs and the spinons,
giving a dual Lagrangian of the form:  ${\cal L}_D = {\cal L}_f + {\cal
L}_v
+{\cal L}_a$ with a vortex piece of the Ginzburg-Landau form,
\begin{equation}
{\cal L}_v(a_\mu) = {\kappa_\mu \over 2} |(\partial_\mu - i2\pi a_\mu )
\Phi |^2
- r |\Phi |^2 - u |\Phi |^4  ,
\label{ldualv}
\end{equation}
and
\begin{equation}
{\cal L}_a = {1 \over {2 \kappa_0}} (e_j^2 - b^2)
+ {1 \over {2\kappa_0}} J_\mu \epsilon_{\mu \nu \lambda} \partial_\nu
a_\lambda  .
\label{lduala}
\end{equation}
Here $e_j = (\partial_j a_0 - \partial_0 a_j)$ and $b=\epsilon_{ij}
\partial_i a_j$ are dual ``electric" and ``magnetic" fields.
The dual magnetic field $b$ is simply the total charge density (in units
of the
Cooper pair charge).
The last term in ${\cal L}_a$ is the only one
coupling the spinons to the vortices.
However, this dual Lagrangian is not valid since it ignores
a strong statistical gauge interaction between spinons and hc/2e
vortices.  To see this consider taking a spinon ($f$) around a closed
loop which encircles an hc/2e vortex.  Along this circuit the
phase $\varphi$
winds by $2 \pi$.  Due to the $1/2$ in Eqn. (\ref{nodondef}) this
implies
that the spinon $f$ must change sign upon completing this circuit.
This can be formally implemented by introducing branch
cuts emanating from each and every vortex, across
which the fermion wavefunction must change sign.
This represents a strong and long-ranged
``statistical" interaction between the spinons and hc/2e vortices.

The presence of this long-ranged interaction clearly invalidates the
form of the dual Lagrangian.  One is tempted to
try and incorporate the branch cuts by introducing
two new gauge fields coupled together by a Chern-Simons
interaction\cite{Lutt}. The most natural way of doing this
is to introduce a coupling ${J_\mu} {\alpha_\mu}$
of the nodons to a gauge field $\alpha_\mu$
which attaches half of a fictitious flux quantum to
each vortex. However, such a coupling is not gauge invariant
since the nodon current, $J_\mu$ is not conserved.
To avoid this problem -- but at the cost of breaking
spin-rotational invariance -- we couple $\alpha_\mu$
to the $z$-component of the nodon spin current,
which is conserved.
Specifically, let $\alpha_\mu$ and $a^s_\mu$
denote statistical gauge fields coupled
to the spinons and vortices, respectively,
with a modified dual Lagrangian:
\begin{equation}
{\cal L}_D = {\cal L}_f + {\cal L}_a + {\cal L}_v(a_\mu + a^s_\mu) +
\alpha_\mu {J_\mu^z} + {\cal L}_{cs}   ,
\end{equation}
and a Chern-Simons interaction:
\begin{equation}
{\cal L}_{cs} =  2 \alpha_\mu \epsilon_{\mu \nu \lambda}
\partial_\nu a^s_\lambda    .
\end{equation}
The Chern-Simons term effectively attaches
flux tubes with strength $1/2$ to each
of the spinons and vortices.
This follows from the equations of motion
obtained from $\partial{\cal L}_D / \partial \alpha_\mu =0$
and $\partial{\cal L}_D / \partial a^s_\mu =0$ which imply,
respectively,
\begin{equation}
\epsilon_{\mu \nu \lambda} \partial_\nu a^s_\lambda = {1 \over 2}
{J_\mu^z}  ,
\end{equation}
and
\begin{equation}
\epsilon_{\mu \nu \lambda} \partial_\nu \alpha_\lambda = {1 \over 2}
j^v_\mu  .
\end{equation}
Here, the vortex three-current is given by,
\begin{equation}
j^v_\mu = {\rm Im}[\Phi^*(\partial_\mu - ia_\mu) \Phi]   .
\end{equation}

Consider now trying to condense the hc/2e vortices.
In the ground state the spinon ($f$-fermions) are gapped out
with $<J^z_\mu> =0$, so one can presumably set $a^s_\mu =0$.
Setting $<\Phi> = \Phi_0$ corresponds to a spontaneous
breaking of the gauge symmetry,
and leads to
an effective Higgs Lagrangian:
${\cal L} = \Phi_0^2 \kappa_\mu a_\mu^2/2$.
In terms of the dual Ginzburg-Landau theory this
describes the ``Meissner state".
But since
the curl of $a_\mu$ corresponds to the total
electrical current, this phase corresponds
to an insulator - the band insulator - with a charge gap.

If the dual Ginzburg-Landau theory is type II
it will exhibit topological excitations
corresponding to penetrating quantized ``dual"
flux tubes.  In the electronic insulator these
correspond to gapped charge $\pm 2e$ spin-zero states,
which are two-electron bound states.

But now consider an excited state in the insulator
which carries spin $1/2$.  This can be created
by adding a spinon at the origin
by acting with $f_\alpha^\dagger(x=0)$.
The presence of a spinon induces a statistical gauge field
from the Chern-Simons term:
\begin{equation}
h_s(x) = \epsilon_{ij} \partial_i a^s_j = {1 \over 2} J_0 = {1 \over 2}
\delta^2(x)   .
\end{equation}
Since $h_s(x)$ corresponds to an applied
``magnetic field" in the dual Ginzburg-Landau theory,
adding a spinon
is equivalent to the insertion
of a solenoid carrying one-half of a (dual) flux quantum.
Being in the Meissner state, the dual Ginzburg-Landau theory
will tend to screen out this applied magnetic field
by generating currents that induce an opposing internal field, $b(x)$.
This follows readily from the energy in the Meissner state
which takes the form:
\begin{equation}
E = { {\Phi_0^2 \kappa_0} \over 2} \int d^2x |\nabla \theta - 2\pi
\vec{a} - 2\pi \vec{a}^s |^2 ,
\end{equation}
where we have put $\Phi = \Phi_0 e^{i \theta}$.  With both $\theta=0$
and
$a_\mu=0$ the energy in the presence of the solenoid will
diverge logarithmically with system size.
Apparently, the energy of an isolated spinon
diverges in the thermodynamic limit.
But in the presence of an induced
internal magnetic field
the energy will be finite provided
the integrated flux is precisely one-half of the dual flux quantum:
\begin{equation}
\int d^2x b(x) = - {1 \over 2}  .
\end{equation}

In physical terms this corresponds to an induced
cloud of electric charge with magnitude $e$.  Evidently, in the
insulating
phase an isolated ${S_z}=1/2$ excitation
will bind charge $e$ to form a
spin-up electron.
The resulting excitation has finite energy.
Similarly, an isolated ${S_z}=-1/2$ will bind charge $-e$ to form a
spin-down hole with the same energy as
the spin-up electron.
This mechanism for confinement of spin and charge
is reminiscent of the confinement of charge and flux
which occurs in the quantum Hall effect.  In the bosonic
Chern-Simons formulation of the quantum Hall effect,
the confinement of charge and flux also
arises via a Higgs mechanism  - in this case when the
composite Boson condenses.

For the above Ginzburg-Landau/Chern-Simons theory
there is another finite energy configuration in the presence of
a single ${S_z}=1/2$ - a $2\pi$
winding in the phase $\theta$ of the vortex field together
with an internal field of one-half quantum which
{\it aligns} with the ``applied flux" ($\int d^2 x b(x) = +1/2$).
This creates a finite energy excitation with
spin one-half and charge $-e$, corresponding
to a conventional spin-up hole. We can create a
spin-down electron of equal energy in a similar manner.
On physical grounds
the energy of the ${S_z}=\pm 1/2$ states should clearly
be degenerate.  Unfortunately, for the above
Ginzburg-Landau/Chern-Simons theory
while both energies
are finite, they will in general be different, due to differences
in the core energies (eg. near the ${S_z}=-1/2$
state  with $\theta$ winding, it is necessary to suppress
the magnitude of $\Phi$ to
zero).   This signals a clear deficiency in the  Chern-Simons
formulation
of the ``statistical" interaction between
spinons and hc/2e vortices.
Since the vortices and spinons can sense the {\it sign}
of the statistical flux (that is $\pm 1/2$ flux quanta
are {\it not} identical) the Chern-Simons fields
do not give a faithful representation of the
branch cuts. The necessary evil of breaking spin-rotational
symmetry is a consequence of this
asymmetry in the Chern-Simons formulation.

Currently, we do not have a convenient formulation
of interacting hc/2e vortices and
spinons which correctly respects this symmetry.
Such a formulation would be particularly desirable
for the case of a d-wave superconductor where the
quasiparticles are gapless.
Nevertheless,
we believe that the Chern-Simons formulation does
help elucidate the correct mechanism behind confinement
of spin and charge upon condensation of hc/2e vortices.

The preceding considerations dovetail naturally
with the following approach to understanding spin-charge
separation. Consider the following {\it gedanken}
experiment, for which we are indebted to B. I. Halperin\cite{gedanken}
, which probes
the existence of spin-charge separation.  Consider a totally
gapped system which
exhibits spin-charge separation, meaning that it has weakly-coupled
neutral spin $1/2$ excitations -- to which we'll
liberally apply the term nodons (even though
the rest of our discussion applies also to systems
which have nothing to do with $d$-wave superconductivity),
following Ref.~\onlinecite{BFN} --
and charge $e$ spin-less excitations -- which we'll call holons,
following Ref.~\onlinecite{krs}. Imagine imposing two
spatially localized perturbations. These
take the form of an interaction Hamiltonian
\begin{eqnarray}
  H_{int} & = & \lambda \left[ \left( Q({\bf x})- e\right)^2 + \left(
      S^z({\bf x})\right)^2 \right] \nonumber \\
  & & + \lambda \left[ \left( Q({\bf x}')\right)^2 + \left(
      S^z({\bf x}')- 1/2\right)^2 \right].
  \label{eq:perturbations}
\end{eqnarray}
Here $Q({\bf x}) = \sum_{\bf y} \rho({\bf x} +{\bf y})
\exp(-|y|^2/\xi^2)$, and $S^z({\bf x}) = \sum_{\bf y} s^z({\bf x}
+{\bf y})\exp(-|y|^2/\xi^2)$ are the total charge and z-component of the
spin
within a smoothed region of linear size $\xi$ around the point ${\bf
  x}$.  The perturbation favors localizing a charge $e$ {\sl without
spin} near ${\bf x}$
and a spin $1/2$ {\sl without charge} near ${\bf x}'$.
Now imagine taking $|{\bf x}-{\bf x'}| \gg \xi \gg a$ (the
lattice spacing), so that the points are well separated.  For small
$\lambda$, the ground state of the system will be unchanged, since
there is a gap to all excitations.  Increasing
$\lambda$ will ultimately induce a change in the ground state to take
advantage of these perturbations.  Provided $\xi$ is taken larger than
the size of the nodon and holon, these excitations can come into the
system to lower its energy and will be localized in the wells.  This
change in the ground state will occur at finite $\lambda$, since the
energy gap to the nodon and holon is finite, and indeed the critical
$\lambda$ will saturate as $|{\bf x}-{\bf x}'| \rightarrow \infty$.
One can interpret this critical $\lambda_c$ as the minimum energy
needed to produce an unbound nodon-holon pair.

Now imagine repeating the same experiment on a band
insulator or any other state which
does not exhibit spin-charge separation.  In this case, the nodon and
holon are not available to ``fill'' the local perturbations.  Instead
the system must create a non-local superposition of elementary
excitations, i.e. develop a polarization, to localize the desired
charge $e$ or spin $1/2$.  Consider for example the region around the
point ${\bf x}$, in which a charge $e$ should localize.  A simple and
generic model for a band insulator is a collection of deep potential
wells,
each containing two electrons.  For the low-energy states, the wells may
be
approximated as quadratic and, in the ground state, all electrons are
in the lowest harmonic oscillator level.  The lowest excited states
which do not induce local spin textures
are constructed simply by moving both electrons in one of the wells from
the ground state to one of the first excited states with  energy gap
$\omega_0$.  A convenient basis for these states is the set $|\mu\rangle
= {a_{\uparrow\mu}^\dagger}
{a_{\downarrow\mu}^\dagger} |0\rangle$, where $a_{\alpha\mu}^\dagger =
\sqrt{m\omega_0/2}(x_\mu-ip_\mu /m\omega)$ is the raising operator along
the $i$ axis in space.  Now consider the superposition of
the ground and excited states,
\begin{equation}
  |{\bf u}\rangle = (1+|u|^2/\ell^2)^{-1/2} \left( |0\rangle +
    {u^\mu \over \ell}  |\mu\rangle \right),
\end{equation}
where $\ell = \sqrt{2/m\omega_0}$ is the characteristic spatial width
of the harmonic oscillator levels.  For small $|u|$, the state $|{\bf
  u}\rangle$ represents a small displacement of the electrons, i.e.
\begin{equation}
  \langle {\bf u}| {\bf x}|{\bf u}\rangle = {\bf u}.
\end{equation}
Thus it possesses a dipole moment ${\bf d} = -2e{\bf u}$, and a local
charge may be built up by polarizing the collection of all the
electrons near the point ${\bf x}$ (i.e. forming a many body state
which is a direct product of single-particle states with differing
displacements on each site).  We consider a slowly-varying
displacement ${\bf u}({\bf x})$ for which a continuum description is
adequate (although this is not a necessary restriction). Far away from
the point ${\bf x}$, the induced charge density $\rho = 2e\bbox{\nabla}
\cdot {\bf u}$, and hence the charge in a given region $R$ is $\int_R \!
d^d{\bf x} \rho({\bf x}) = 2e \int_{\partial R} \! {\bf u}\cdot
d{\bf\hat{n}}$.  The radially symmetric configuration
\begin{equation}
  {\bf u}({\bf x''}) = {1 \over {2S_d}} {{\bf x-x''} \over {|{\bf
x-x''}|^d}}
\end{equation}
thus carries a total charge $e$.  Provided all the electrons are
involved in the texture, there is no net spin polarization.  This
polarized state is not an eigenstate of the unperturbed Hamiltonian,
but does couple favorably to the first term in
Eq.~\ref{eq:perturbations}.  We can however determine the expectation
value of the (unperturbed) energy in this state.  The result is
essentially classical: $E({\bf u}) = \int\! d^d{\bf x} \omega_0
(u/\ell)^2$.  In two dimensions, this integral is logarithmically
divergent, $E({\bf u}) \sim \ln L$.  The isolated long-range
polarization thus cost an {\sl infinite} energy.  In the thought
experiment, this divergence will be cut off by the finite distance
between ${\bf x}$ and ${\bf x'}$, since we may localize an oppositely
charged texture around the point ${\bf x'}$ in combination with an
added electron with $S^z = 1/2$, thus satisfying both perturbations
and rendering the energy finite.  However, the critical $\lambda_c$
will grow logarithmically as $|{\bf x-x'}| \rightarrow \infty$, and
hence it becomes impossible to create the isolated holon and nodon in
the thermodynamic limit.

This argument is appealing in
that it agrees with the earlier Chern-Simons calculations which
suggested logarithmic confinement of holons and spinons.  As a means
of distinguishing spin-charge separated from spin-charge
confined phases, however, it is somewhat
delicate.  In particular, it fails for $d>2$, where the polarization
energy to create the charge $e$ texture becomes finite.  It is also
somewhat unsatisfying because the texture is not an eigenstate of
the unperturbed Hamiltonian. Fortunately, the duality formalism allows
the two phases to be distinguished instead by the dual order parameter
$\Phi$. When $\Phi$ condenses, the statistical gauge interactions
between nodons and the condensing vortices leads to spin-charge
confinement in $d=2$. In this sense $\Phi$ is an
{\sl order parameter for confinement}, a
rather unique feature of the present theory.

\section{Mott Insulators, $hc/e$ Condensates, and the Nodal Liquid}

We now turn to the more interesting case of insulating phases with one
electron per unit cell.  In such Mott insulators electron interactions
are necessary to destroy the metallic state, in contrast to the band
insulator which has a smooth non-interacting limit. As in Section II
we will obtain a description of the insulating state by quantum
disordering an appropriate superconducting phase.  A brief discussion
of quantum-disordered s-wave superconductivity at this density
illustrates the need to consider ``double strength'' vortex
condensates in the cuprates.  Such $hc/e$ condensates are {\sl nodal
Liquid} insulators, and the subject of the remainder of this section.

\subsection{Quantum-disordered s-wave}

We begin by considering a system of spinful interacting electrons
moving in the 2d continuum, which pair to form a spin singlet s-wave
superconducting phase, and then ``turn on" a periodic potential which
for simplicity has square symmetry.  Here, however, we choose the period
to
correspond to one electron per unit cell.  As in Section II, with
increasing potential strength the superconducting phase can be
destroyed by the unbinding and condensation of $hc/2e$ superconducting
vortices.  But in this case there is only one-half of a Cooper pair
per unit cell, so the resulting insulating phase will be dramatically
different from the band insulator.  In particular, one expects the
formation of a crystalline state of Cooper pairs which exhibits
charge-density-wave ordering and spontaneously breaks (discrete)
translational symmetry.

In order to proceed expeditiously to our main
interest, the nodal liquid, we only mention a few salient points here.
The difference between the CDW and the band insulator
in the present approach is that the Berry's
phase term\cite{Duality},
\begin{equation}
  {\cal L}_{Berry} =  n_0 \partial_t \varphi,
  \label{lberry}
\end{equation}
cannot be dropped. To appreciate the physics of this
term, we must return to the lattice, where the
lattice Hamiltonian corresponding to Eq.~\ref{lphi},
together with the Berry's phase term (Eq.~\ref{lberry}),
takes the form:
\begin{equation}
  {\cal H}_\varphi = -t_2 \sum_{\langle i,j\rangle} \cos(\varphi_i -
\varphi_j) +
  U_2 \sum_i (n_i - n_0)^2  .
\end{equation}
Here, $n_i$ is the Cooper pair number operator
which is canonically conjugate to $\varphi$.
Because, at half-filling, there is, on average, half
a Cooper pair per site, one has $n_0 = 1/2$.

Implementing duality as before\cite{Duality} (but now
on the lattice) gives a dual Euclidean action
which is the lattice analogue of Eq.~\ref{ldualv}\ and
Eq.~\ref{lduala}, the only
difference being that the Maxwell term, ${1 \over {2 \kappa_0}}(e_j^2 -
b^2)$,
term is replaced by:
\begin{equation}
  {\cal S}_a = {u_2 \over 2} \sum_{j\mu} (\epsilon_{\mu\nu\lambda}
  \Delta_\nu a_j^\lambda - n_0
  \delta_{\mu 0})^2    .
\end{equation}

The new feature, as compared to the last section,
is the ``off-set" charge $n_0$ which results from
the Berry phase term (\ref{lberry}).
It corresponds to an applied ``magnetic" field for
the lattice Ginzburg-Landau theory.
The insulating CDW phase of the Cooper pairs
corresponds to an Abrikosov flux lattice
in this dual representation. As in the case of the band insulator,
the vortex condensate leads to a charge gap and insulating
behavior.

Generally, one expects that a condensation of $hc/2e$ vortices
will lead to charge ordering with charge $2e$ per unit cell.
This follows from the underlying duality:  $hc/2e$
vortices pick up a $2\pi$ phase change
upon encircling charge $2e$ Cooper pairs -- the same phase
accumulated when a Cooper pair encircles such a vortex.
Thus, the liberation and condensation
of $hc/2e$ vortices leads to a charge
quantization in units of $2e$.  For the 
model with attractive interactions and one electron per unit cell
considered above,
the resulting state
is the (charge $2e$) Cooper pair crystal.

\subsection{Quantum-disordered d-wave}

We now turn to the interesting problem of the quantum-disordered {\sl
d-wave} superconductor.  D-wave superconductivity is thought to
arise from the combination of strong on-site Coulomb repulsion and
some unspecified (and controversial) longer-range attraction on the
scale of a few lattice spacings.  Certainly strong local repulsion is
a key ingredient of the cuprates.  For such systems,  the $hc/2e$ vortex
condensation
described above -- which implies
considerable double occupancy (at least in a region near
half-filling) -- is physically unreasonable.
It must also be discarded on phenomenological grounds,
as the actual undoped  antiferromagnet is a Mott insulator
without charge ordering.

Although $hc/2e$-flux vortex unbinding is untenable for this case,
phase coherence must nevertheless still be destroyed to obtain an
insulating state.  Charge uniformity and phase disruption can both be
achieved together by unbinding bound {\sl pairs} of vortices with flux
$hc/e$ instead of isolated ones.  We expect such a {\sl
 double-vortex condensate} to appear in the dual description as a
condensate with a doubled dual charge, and hence a halved dual flux
quantum\cite{sachdev}.  At half-filling then, the dual lattice Ginzburg-Landau
theory has a full $2\pi$ flux per plaquette and thus, as desired,
exhibits no translational symmetry breaking.

Having motivated double-vortex condensation in the d-wave case, we now
proceed to discuss its implementation.  The calculations are
significantly different because of an important additional physical
ingredient in the d-wave superconductor: gapless fermionic
quasiparticles.  These excitations arise owing to the vanishing of the
amplitude of the pair wavefunction at its nodes in momentum space.
The presence of low-energy fermionic excitations necessitates a
careful reinvestigation of the duality transformation and its
implications.  Much of the necessary
calculations and formalism was discussed in detail in
Ref.~\onlinecite{BFN}, and is briefly recapitulated in
the appendix.

As for the s-wave case, the analysis of the interactions of vortices
with quasiparticles is based on the neutralizing change of variables
in Eq.~2.4.  The distinctive feature of the d-wave superconductor is
that the neutral spin-$1/2$ particles are {\sl gapless}, and can be
described by a Dirac Hamiltonian.  Because in this case the spinons
near the nodes can contribute to low-energy physics, we attribute to
them special significance and the name {\sl nodons}, signifying the
low-energy spinons descended from the d-wave nodal quasiparticles.
Being gapless, they can be described by continuum field theory and a
4-component Dirac spinor $\psi$ (the analog of $f$ in Sec.~II -- see
the appendix for a precise definition).

Having already argued that only $hc/e$-flux vortices should be
considered, we will focus primarily on this simpler case.  It is
however, appropriate at this point to reflect briefly on the
consequences of these strong gauge interactions should single-strength
vortices become important low-energy excitations.  As above,
$hc/2e$-vortex unbinding gives rise to strong gauge interactions of
the spinons, now nodons.  The gauge-theoretical arguments given in
Sec.~II can be again carried through, and we expect confinement of the
nodons.  Unlike the s-wave case, however, because the nodons are
gapless excitations, their presence or absence has definite
consequences on the {\sl ground state} correlations.  For instance, in
a pure d-wave superconductor, the gapless Dirac excitations lead to
static (equal time) power-law spin correlations and a $T$-linear
magnetic susceptibility.  It is natural to expect that the removal of
the nodons from the low-energy spectrum in the insulator will be
accompanied by the condensation of some pairing operator (for
example, spin-density-wave order is characterized by the order
parameter $\langle \psi^\dagger \tau^y
\sigma^y\vec{\sigma}\psi^\dagger \rangle \neq 0$; other possibilities
are legion), in most cases accompanied by a gap for the reconfined
electrons.\cite{foot2}\ The formation of such a paired-nodon
state is analogous to {\sl chiral symmetry breaking} in QCD, and in
that context as well is generally believed to accompany confinement
(although the converse need not be true).  One can imagine approaching
such a state from the superconductor by continuously lowering the
vortex ``mass'' to zero, at which point one has a theory in which
gapless fermionic nodons interact with gapless bosonic vortices via
strong gauge interactions.  This putative critical point is a tempting
starting point for future systematic studies of such instabilities.

We now return to the problem of $hc/e$ vortex condensation, which,
although motivated on energetic grounds, has dramatic consequences for
the elementary excitations.  Examining again Eq.~2.4\ when only
double-strength vortices are present, $\varphi$ is defined modulo
$4\pi$, and this transformation defines a single-valued neutral
fermion.  The nodons thus experience no statistical gauge interactions
in this case.  The results obtained in Ref.~\onlinecite{BFN}, which
ignored gauge interactions, hence apply to the $hc/e$ condensate,
with the proviso that the fundamental vorticity must be doubled
throughout the analysis. The salient result is that when gauge effects
are absent, the nodons and vortices interact only via the two-fluid
interaction Lagrangian
\begin{equation}
  {\cal L}_{int} = \partial_\mu \varphi J_\mu ,
  \label{eq:Lint}
\end{equation}
where $J_\mu$ is the electrical current carried by the quasiparticles,
and is bilinear in the $\psi$ fields.  Eq.~\ref{eq:Lint}\ can be
understood as the Doppler shift of the nodon energies in a superflow
given by $\partial_\mu\varphi$.  This is a much weaker coupling than
the statistical gauge interactions in the single-vortex condensate,
and controlled analytic calculations are possible.  Detailed
predictions for this quantum-disordered state, the {\sl nodal Liquid}
(NL), can be derived by writing a coarse-grained continuum theory for
$\Phi_2$ and $a_\mu$, as in Refs.~\onlinecite{BFN,Fisher98}.  The key
conclusions are: (1) gapless nodons survive into the NL state,
carrying spin but neither charge nor current at low frequencies; (2)
the NL has gapped charged excitations, the lowest-lying of which are
expected to be (spinless) charge $\pm e$ holons, which occur as vortices in the
dual order parameter $\Phi_2$; (3) the half-filled NL has uniform
charge density, and upon (hole) doping charge is introduced as a
spin-less Wigner crystal with charge $e$ per unit cell (but see Sec.~IV
for a discussion of how this may be modified when antiferromagnetism
is present).

It is important to emphasize that a connection has been made here
between two apparently unrelated phenomena.  By {\sl assuming}
double-vortex condensation, characterized by the dual order parameters
$\Phi_2 = \langle e^{2i\theta_j}\rangle \neq 0$ (double vortices are
condensed), $\Phi = \langle e^{i\theta_j}\rangle =0$ (single vortices
are bound), we were led to the persistence of spin-charge deconfinement
in
the insulator.  The single-vortex disorder parameter, $\Phi$ (which
also distinguishes translational symmetry breaking at
half-filling) can thus be regarded
as an {\sl order parameter for confinement}.

\subsection{Lattice model for the nodal liquid}

We conclude this section by describing a
direct route to the NL at half-filling, by which most of its
properties may be derived without the use of the duality mapping.  To
do so, consider the following lattice regularization, which forbids
$hc/2e$-flux vortices from the outset:
\begin{eqnarray}
  H_{qp} & = & \sum_{\langle ij\rangle} -t\left( c_i^\dagger
    c_j^{\vphantom\dagger} + c_j^\dagger
    c_i^{\vphantom\dagger}\right)  \nonumber \\
  & & + |\Delta|(-1)^{x_i-x_j}
  e^{-i(\varphi_i+\varphi_j)/2} c_{i\uparrow}^\dagger
  c_{j\downarrow}^\dagger + {\rm h.c.}, \\
  H_\varphi & = & \sum_{\langle ij\rangle} -J \cos\left( {\varphi_i
      \over 2} - {\varphi_j \over 2}\right) \nonumber \\
  & & + \sum_i {U \over 2}
  \left(2 n_i + c_{i\alpha}^\dagger c_{i\alpha}^{\vphantom\dagger} -
1\right)^2.
\end{eqnarray}
The $\cos(\varphi_i/2-\varphi_j/2)$ term has been chosen to allow $\pm
4\pi$ phase slips but not $\pm 2\pi$ phase slips, hence ``confining''
single-strength vortices.  Further, we have made the apparently
arbitrary choice of dividing the superconducting pair-field phase
amongst neighboring sites.  While this may seem unnatural, provided
the continuum d-wave quasiparticle Hamiltonian is an adequate
low-energy description, {\sl any} lattice
regularization should reproduce identical low-wavelength behavior.
Finally, we have included a ``charging energy'' term coupling to the
total (Cooper pair + quasiparticle) charge.

This model has particle-hole symmetry, and at zero chemical potential
is thus automatically at half-filling.  To determine the properties of
the system, we begin by performing the lattice analog of
Eq.~\ref{nodondef}: $c_{j\alpha}^\dagger = e^{i\varphi_j/2}
f_{i\alpha}^\dagger$.   Simultaneously, to avoid non-trivial
commutation relations between $f,f^\dagger$ and $n$, we let $N_i = 2
n_i + c_i^\dagger c_i^{\vphantom\dagger}$ and $\phi_i = \varphi_i/2$.
The $f$ fermion creates neutral, spin-$1/2$ quanta.  In these
variables, the Hamiltonian becomes $H=H_\varphi+H_f+H_{int}$, with
\begin{eqnarray}
  H_\phi & = & \sum_{\langle ij\rangle} -J \cos\left( \phi_i -
      \phi_j \right) + \sum_i {U \over 2}
  (N_i-1)^2, \\
  H_f & = & \sum_{\langle ij\rangle} |\Delta|(-1)^{x_i-x_j}
 f_{i\uparrow}^\dagger  f_{j\downarrow}^\dagger + {\rm h.c.}, \\
 H_{int} & = & \sum_{\langle ij\rangle} -t \left(
   e^{i(\phi_i-\phi_j)} f_{i\alpha}^\dagger
   f_{j\alpha}^{\vphantom\dagger} + {\rm h.c.}\right).
\end{eqnarray}
Note that the nodon-phase coupling has been transferred from the
pair-field interaction to the kinetic term by the operator
transformation.  An insulating state is obtained in the limit $U \gg
J,t$, where the charging energy dominates over both pair and
single-particle hopping.  This state can be studied perturbatively in
$t$ and $J$, expanding around the insulating state with $N_i=1$ {\sl
  exactly} on each site.  At $t=J=0$, however, the
$f$-particle Hamiltonian is still highly degenerate.  This
degeneracy is broken at second-order in $t$ and $J$, giving the
effective Hamiltonian (obtained, e.g., by perturbatively integrating
out $N_i$ and $\phi_i$ in a path-integral formulation)  $H_{eff} =
H_{e0}
+ H_{e1}$ with
\begin{eqnarray}
  H_{e0} & = & \sum_{\langle ij\rangle} - {tJ \over {U}} \left(
    f_{i\alpha}^\dagger
    f_{j\alpha}^{\vphantom\dagger} + {\rm h.c.}\right) \nonumber \\
  & & + \sum_{\langle ij\rangle} |\Delta|(-1)^{x_i-x_j}
  c_{i\uparrow}^\dagger  c_{j\downarrow}^\dagger + {\rm h.c.},
  \\
  H_{e1} & = & - {2t^2 \over U} \sum_{\langle ij\rangle}
  f_{i\alpha}^\dagger
  f_{j\alpha}^{\vphantom\dagger} f_{j\beta}^\dagger
  f_{i\beta}^{\vphantom\dagger} .
  \label{eq:He1}
\end{eqnarray}
The quadratic Hamiltonian $H_{e0}$ is identical to the mean-field
Hamiltonian for d-wave quasiparticles, with a renormalized bandwidth
$8tJ/U$.  It therefore describes two sets of spin-$1/2$ Dirac fermions
at low energies.  One thus recovers in this way the NL phase obtained
previously via continuum duality.  The
interaction $H_{e1}$ can be rewritten as a combination of
antiferromagnetic exchange and contact repulsion of the $f$-particles.
If both are weak (as in the large $U$ limit), such four-fermion
interactions are strongly irrelevant around the non-interacting NL
fixed point, due to the linearly vanishing density of states of the
Dirac fermions.  A slightly refined analysis including a physical
external gauge field $A_\mu$ allows one to calculate explicitly the
conductivity and show that the $f$-fermions carry no current in the NL
state, so it is indeed an insulator\cite{BFN}.

\section{Antiferromagnetism}

\subsection{Phenomenology}

As discussed in the previous section, $hc/e$-flux vortex condensation
atop the d-wave superconductor yields the NL, an insulator with charge
quantization (charge $e$ per unit cell) appropriate near the
half-filled Mott insulator.  Unlike the CDW obtained by $hc/2e$-flux
vortex condensation or the conventional (fully gapped ``short-range
RVB'') spin liquid state,\cite{krs}\ the NL also contains low-lying
gapless spin
degrees of freedom, the nodons, which contain the germ of true
antiferromagnetic order.  As described in Ref.~\onlinecite{BFN}, the
phenomenology described above can be easily extended to include N\'eel
order.

The description to this point has essentially neglected inter-nodon
interactions.  This approach is justified provided such interactions
are weak, as all such terms are perturbatively irrelevant (in the
renormalization group sense) in the Dirac theory describing the
NL.\cite{BFN}\ However, perturbative irrelevance does not imply that
{\sl strong} interactions cannot drive quantum phase transitions and
hence a qualitative change in behavior.  Indeed, interacting Dirac
fermion models are known to undergo {\sl chiral symmetry breaking}
transitions as quartic couplings are increased.\cite{Polyakov87}\ The
nature of the transition incurred depends upon the precise nature of
the interactions, and various circumstances can induce
antiferromagnetism, spin-Peierls, charge density wave, and other types
of ordering from the NL Lagrangian.  Indeed, in the lattice model
above we obtained an interaction (Eq.~\ref{eq:He1}) capable of driving
a transition to an antiferromagnetic state if sufficiently large.

Because of the uncertainties and pitfalls of attempting a microscopic
justification of such interactions, however, we prefer to follow the
strategy of Ref.~\onlinecite{BFN}\ and take a more phenomenological
approach.  For simplicity let us focus on the case of half-filling
with particle/hole symmetry.  Since the cuprate materials are clearly
antiferromagnetically ordered, we will {\sl assume} the existence of a
triplet collective mode with momentum $(\pi,\pi)$.  In the NL phase
this magnon mode has a gap, and indeed we expect also a non-zero
lifetime, as the triplet magnons can decay into pairs of nodons, so
strictly speaking the magnons are not sharply defined elementary
excitations in this case.  Nevertheless, we may imagine tuning a
parameter (e.g. reducing frustrating spin-spin interactions in a
lattice model) to reduce the magnon gap.  Ultimately when it vanishes
the collective mode becomes sharp and condenses to form an
antiferromagnetically ordered state.  In the antiferromagnetic state,
the non-zero N\'eel vector coherently mixes nodons with opposite
quasi-momentum and opposite spin, halving the magnetic Brillouin zone
as is usual in spin-density-wave systems.  As in those more
conventional cases, this has the effect of opening up a gap in the
nodon spectrum.
\begin{equation}
  E_n(\vec{q}) = \pm \sqrt{ (v_F q_\perp)^2 +
(v_{\scriptscriptstyle\Delta} q_\parallel)^2 +
    (gN_0)^2 }  ,
  \label{eq:nodongap}
\end{equation}
where $v_F$ and $v_\Delta$ are the Dirac ``velocities'' perpendicular
and parallel to the putative Fermi surface, ${\bf q}$ is the momentum
measured from a node, $N_0$ is the mean-field staggered magnetization,
and $g$ is a phenomenological coupling constant.  Due to this mixing,
the only gapless degrees of freedom in the antiferromagnet are the
collective spin-wave modes guaranteed by Goldstone's theorem.

In this way we arrive at an effective low-energy field theory for an
antiferromagnetic Mott insulator, which we will denote (for reasons
which will become apparent) as an AF* phase.  The antiferromagnetic
spin order, featureless incompressible charge configuration, and
gapless spin waves are qualitatively identical to those we would
obtain in more conventional antiferromagnetic models, e.g. the nested
spin-density-wave in the weakly-interacting half-filled Hubbard model,
or alternatively the $t-J$-like very large $U$ limit of the same
Hamiltonian.  We stress, however, that although the nodons have been
lifted away from zero energy, they are not confined by the spontaneous
symmetry breaking in the AF* phase, i.e. spin-$1/2$ neutral particles
still exist as well-defined elementary excitations.  For this reason,
we believe that such a novel antiferromagnetic insulator (AF*) is
topologically distinct from (i.e. cannot be
adiabatically deformed into) a more conventional antiferromagnetic
(AF) state.  This conviction is bolstered by the existence of the dual
order parameter $\Phi_2$, which we have argued characterizes
the nodal liquid and the AF* phase.  Since $\Phi_2$
creates $hc/e$ vortices in a {\it pair} field,
the AF* phase contains the germ of superconductivity.
In contrast,
construction of a dual order parameter
for the conventional antiferromagnet
probably requires the use of Chern-Simons
(charge $e$) bosons, obtained directly by statistical transmutation
from the electrons.  For example,
condensing
elementary $hc/e$ vortices in the spin-up boson
to form a charge $e$ crystal which lives
on one sub-lattice, and similarly freezing the
spin down particles onto the other sub-lattice,
should suffice to describe
the hidden order of the AF phase.
Evidently, the dual order parameters in AF and AF*
are very different.
Alternatively, one can distinguish AF* and AF by
testing for the presence or absence of spin-charge separation
employing the argument in Section II.  To make the argument precise
in this case probably requires adding an easy-axis anisotropy,
\begin{equation}
  H_{ea} = \sum_{\bf x} J_{ea}\left[ (s^x)^2 + (s^y)^2 \right].
  \label{eq:easyaxis}
\end{equation}
which creates a gap in the magnon spectrum.
But it is of more interest
to address whether the AF and AF* phases
can be
distinguished {\it experimentally}?  To address this,
we turn to a discussion of the electron spectral
function in these two phases.

\subsection{electron spectral function}

In standard many-body systems, the existence of well-defined
excitations is ascertained by
examination of the relevant spectral function.\cite{Negele88}\
Unfortunately, a direct probe
of spin-charge separation via spectral functions
is not possible, since
there are no
local operators which separately create nodons and holons.
On the other hand, the electron spectral
function, $A({\bf k},\omega)$, is accessible experimentally
and has been intensively studied
in the high-temperature superconductors
with momentum resolution
via angle-resolved photo-emission spectroscopy\cite{Loeser,Norman,Shen}\
and locally (i.e. in
momentum-integrated form) via non-linear tunneling
characteristics.\cite{Miyakawa}\
It seems natural to suggest
that $A({\bf k},\omega)$ might possibly give one a way to distinguish
the AF and AF* phases, since in the latter unbound
nodon-holon pairs form a two-particle continuum, and in the former the
electron is itself the elementary excitation.

\subsubsection{AF}

To see how this idea works out in practice, let us first consider in
some detail the spectral function in the AF phase.  A simple model
which captures the qualitative physics of the spectral function is
fluctuation-corrected spin-density-wave mean field theory.  The
(imaginary time) quasiparticle Lagrangian is
%\begin{eqnarray}
%  L & = & \int_{\bf k} c_{\bf k}^\dagger \left[ \partial_\tau +
% \epsilon_k
%    - \mu
%    \right]c_{\bf k}^{\vphantom\dagger} \nonumber \\
%    & & + \tilde{g} \vec{N} \cdot
%    c_{\bf k + Q}^\dagger \vec{\sigma} c_{\bf k}^{\vphantom\dagger},
%\end{eqnarray}
\begin{equation}
  L  =  \int_{\bf k} c_{\bf k}^\dagger \left[ \partial_\tau + \epsilon_k
- \mu \right]c_{\bf k}^{\vphantom\dagger}
  + \tilde{g} \vec{N} \cdot
  c_{\bf k + Q}^\dagger \vec{\sigma} c_{\bf k}^{\vphantom\dagger},
\end{equation}
where we have assumed ordering at ${\bf Q} = (\pi,\pi)$.  For
simplicity, let us assume the Fermi surface intersects $(\pi/2,\pi/2)$
with some curvature (it is straightforward to generalize this to other
geometries).  Choosing new coordinates along the $(1,1)$ and $(1,-1)$
axes, we then write ${\bf k} = (\pi/\sqrt{2},0) + {\bf q}$, and
$\epsilon_{\bf k} -\mu \approx v_F q_x + q_y^2/2m$ near this point.
Performing a similar expansion near the opposite point on the Fermi
surface, and defining continuum fields $\tilde\eta_{\pm}(\bf q) \equiv
c_{(\pm\pi/\sqrt{2},0) + {\bf q}}$, one finds
\begin{eqnarray}
  {\cal L}_{Nqp} & = & \tilde\eta^\dagger \left[\partial_\tau + iv_F
\tau^z
    \partial_x - {1 \over 2m}\partial_y^2\right]
  \tilde\eta^{\vphantom\dagger} \nonumber \\
  & & + g \vec{N} \cdot \tilde\eta^\dagger \vec{\sigma}\tau^x
  \tilde\eta^{\vphantom\dagger},
\end{eqnarray}
where $g$ is the spin-density-wave coupling constant.  We suppress the
additional quasiparticles located near $(\pm \pi/2,-\pi/2)$, since
these are not coupled to the $\eta$ fields by the ordering wavevector
${\bf Q}$.  In the antiferromagnetic phase, $<\vec{N}> = N_0 \ne 0$,
and if fluctuations are ignored the electron states are gapped with
energy dispersion,
\begin{equation}
E({\bf q}) =
\sqrt{|\Delta|^2 + v_F^2 q_x^2} +
q_y^2/2m  ,
\end{equation}
with $\Delta = gN_0$ the mean field spin-density wave gap.

Spatial and temporal fluctuations of the
N\'eel field $\vec{N}$ can be described by, e.g. a
Landau theory such as Eq.~\ref{Lneel}, or by a nonlinear sigma model.
In the AF phase, we require only the spin-wave expansion for small
deviations, $\Pi_i \ll 1$, from perfect alignment, i.e. $\vec{N} = N_0
(\Pi_1,\Pi_2,\sqrt{1-\Pi^2})$, for small $\vec{\Pi}$.  Since uniform
rotations of $\vec{N}$ are equivalent by SU(2) invariance, it is
convenient to perform the ``gauge'' rotation
\begin{equation}
  \tilde\eta({\bf x},\tau) = \exp[i\epsilon_{ij} \sigma_i \Pi_j({\bf
    x},\tau)] \eta({\bf x},\tau).
  \label{eq:su2rot}
\end{equation}
In the new variables, the quasiparticle Lagrangian becomes
${\cal L}_{Nqp} =
{\cal L}_\eta + {\cal L}_{\Pi-\eta}$, with
\begin{eqnarray}
  {\cal L}_\eta & = & \eta^\dagger\left[\hat{l}+ \Delta\sigma^z
    \tau^x\right]\eta^{\vphantom\dagger} , \label{eq:Leta} \\
  {\cal L}_{\Pi-\eta} & = & -i\epsilon_{ij}\eta^\dagger\left[
\hat{l}\Pi_i -
    {1 \over m}\partial_y\Pi_i \partial_y\right]\eta^{\vphantom\dagger}.
  \label{eq:LetaPi}
\end{eqnarray}
Here $\hat{l} \equiv \partial_\tau + iv_F\tau^z\partial_x - {1\over
2m}\partial_y^2$.
Finally, the magnons are governed by the
quadratic Lagrangian
\begin{equation}
  {\cal L}_\Pi = {K \over 2}\left[|\partial_\tau\vec{\Pi}|^2 + v_s^2
    |\nabla\vec{\Pi}|^2\right].
  \label{eq:LPi}
\end{equation}
The spin-waves in Eq.~\ref{eq:LPi}\ are gapless as required by
Goldstone's theorem.  Neglecting the coupling to the spin-waves (a
good approximation if $K$ is very large, so that the $\vec{\Pi}$
fields fluctuate very little), the $\eta$ particles are
non-interacting quasiparticles with a gap $\Delta$, and have a sharp
spectral function
\begin{equation}
A^0_\eta({\bf q},\omega) = \pi^{-1} {\rm Im}
G^0_\eta({\bf q},i\omega_n \rightarrow \omega + i\delta) = \delta(\omega
- E({\bf q}))  .
\end{equation}
The spin-wave
coupling, Eq.~\ref{eq:LetaPi}, generates a self-energy in the $\eta$
Green's function.  The relevant diagram is indicated in Fig.~1, and a
straightforward if tedious evaluation shows that
\begin{eqnarray}
  G_\eta({\bf q},\omega) & = &  [i\omega + v_F \tau^z q_x +q_y^2/2m
  \nonumber \\
  & & +  \Delta\sigma^z
    \tau^x +  \Sigma({\bf q},\omega)]^{-1}, \\
  {\rm Im} \Sigma({\bf q},\omega) & \sim & {1 \over Kv_s^{d+1}}
  (\delta\omega)^d \Theta(\delta \omega), \label{eq:irrelevant}
\end{eqnarray}
with $\delta\omega = \omega - E({\bf q})$. Eq.~\ref{eq:irrelevant}\
holds provided $|q| < m v_s$.
Since ${\rm Im} \Sigma \ll \delta\omega$ for small $\delta\omega$, the
decay rate is negligible at low energies and we expect a
delta-function singularity to survive in the spectral function at
$\delta\omega = 0$ (there will also be some small shifts in the energy
spectrum itself given by the real part of $\Sigma$).
\begin{figure}[htb]
  \hspace{0.5in}\epsfxsize=2.5in\epsfbox{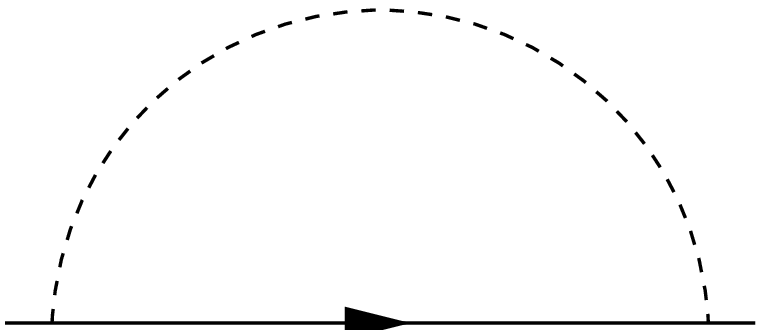}

  {Fig.~1: Self-energy diagram for the charged quasiparticles in the
    AF phase.  The solid line indicates a dressed $\eta$-fermion
    Green's function, while the dashed line is the spin-wave
propagator.}
\end{figure}
When calculating the {\sl electron} spectral function, one needs to
include the effects of the SU(2) rotation, Eq.~\ref{eq:su2rot}.  Since
the physical electrons are created by the $\tilde\eta$ fields,
additional factors of the $\vec{\Pi}$ operators appear in $A({\bf
  k},\omega)$.  The $\Pi$ operators may be expanded out of the
exponential and treated perturbatively.  Both this effect and the
broadening due to the non-zero self-energy above lead to additional
weight for $\delta\omega>0$, which is often referred to as
``incoherent'' spectral weight.  The general expectation for the
electron spectral function in the AF phase is illustrated in Fig.~1.

The physical meaning of these results is the following.  The minimum
energy excitation with charge $e$ and spin-$1/2$ is the electron,
which in the interacting system is ``dressed'' by magnon excitations
that mix with the bare electron.  Further, there are higher-energy
states involving a dressed electron and unbound excited magnons which
are orthogonal to the interacting electron but not the bare one, and
thus show up as continua for $\delta\omega>0$ once interactions are
present.  The true elementary excitation does not decay however,
basically because phase space (which leads to the ${\rm Im}\Sigma \sim
(\delta\omega)^d$ law above) prevents it.  Thus the expected electron
spectral function has a resolution-limited dispersing peak at the
single-particle gap near its minimum, above which lies continuous
spectral weight.  Well away from $(\pi/2,\pi/2)$, phase space may (or
may not) open up to allow decay even of the single-particle peak,
depending upon details of the band structure and interactions.

\subsubsection{AF*}

In the unconventional antiferromagnet, we expect the presence of
unconfined nodons and holons to lead to a two-particle continuum in
the electron spectral function.  While this is indeed the case,
lower-energy features in fact exist due to {\sl nodon-holon bound
states}.  Such bound states are analogous to excitons in
semiconductors, which provide sharp peaks in the optical conductivity
despite the existence of an electron-hole continuum at higher
energies.

Similar considerations apply here.  In particular, if we consider the
interaction of a charge $e$ holon with a spin-$1/2$ nodon, it is quite
natural to expect that they may experience an attractive interaction
leading to a bound state with {\sl both} charge $e$ and spin-$1/2$,
i.e. an {\sl electron}.
To show the existence of such bound states we specialize again
to the case of half-filling with particle-hole symmetry.
It is convenient to perform a particle-hole (Bogoliubov) transformation
on the nodon operators
near one pair of nodes,
\begin{equation}
\tilde{\psi}_{a\uparrow} = \psi_{1a\uparrow}  ;  \hskip0.5cm
\tilde{\psi}_{a\downarrow} = \psi^\dagger_{1a\downarrow}   ,
\end{equation}
so that the number operator
of the transformed fermions is proportional to
the z-component of spin:  $S_z = (1/2)\int d^2 x \tilde{\psi}^\dagger
\tilde{\psi}$.  In the
presence of Neel order, $\vec{N}= N_0 \hat{z}$, the transformed
nodon Hamiltonian density
takes the simple form,
\begin{equation}
{\cal H}_n = \tilde{\psi}^\dagger H_n \tilde{\psi}   ,
\end{equation}
with a single particle Hamiltonian:
\begin{equation}
H_n({\bf p}_1) = v(\tau_z p_1^x + \tau_x p_1^y ) + \Delta_n \tau^y
\sigma^y  .
\end{equation}
Here, the nodon momentum operator
${\bf p}_1 = -i {\bf \nabla}_{r_1}$ is conjugate
to the position ${\bf r}_1$,
and for simplicity we have assumed
only a single nodon velocity.  This Hamiltonian
describes massive nodon states, with energy gap
$\Delta_n = g N_0$.
Since the holons are also gapped,
the appropriate first quantized
Hamiltonian for a single holon (with position ${\bf r}_2$ and momentum
${\bf p}_2$) is simply
\begin{equation}
H_h({\bf p}_2)  = \Delta_h + {{\bf p}^2 \over 2 m_h }   .
\end{equation}

The form of the interaction between the nodons and holons
follows from the dual Lagrangian in Section 2.
For simplicity we only retain the density-density
interaction term, proportional to
$J_0 \epsilon_{ij} \partial_i a_j$, where
$\epsilon_{ij} \partial_i a_j$ is the holon density
and the nodon density can be expressed
in terms of the transformed fermions as:  $J_0 = \tilde{\psi}^\dagger
\tau^z \sigma^z \tilde{\psi}$.  The corresponding first quantized
interaction Hamiltonian is then,
\begin{equation}
H_{int} ({\bf r}_1 - {\bf r}_2) = u a^2 \tau^z \sigma^z
\delta^{(2)} ({\bf r}_1 - {\bf r}_2)  ,
\end{equation}
with interaction strength $u$, and $a$ a short distance cutoff.

Since the two-body Hamiltonian is independent
of the ``center of mass"
coordinate ${\bf R} = ({\bf r}_1 + {\bf r}_2)/2$,
the total momentum, ${\bf P} = {\bf p}_1 + {\bf p}_2$,
is conserved.  For simplicity we consider
bound states with ${\bf P}=0$.
The Hamiltonian for the {\it relative}
coordinates,
\begin{equation}
{\bf r} = {\bf r}_1 - {\bf r}_2  ;  \hskip0.5cm
{\bf p} = ({\bf p}_1 + {\bf p}_2)/2  ,
\end{equation}
then takes the simple form
\begin{equation}
H_{rel} = H_n({\bf p}) + H_h({\bf p}) + H_{int}({\bf r})  .
\end{equation}
To solve for bound states with energy $E$,
we recast the Schr\"odinger
equation $H_{rel} \phi = E \phi$ in the form:
\begin{equation}
G^{-1}({\bf q}) \phi({\bf q}) = -ua^2 \tau^z\sigma^z \phi({\bf r} = 0)
,
\end{equation}
with matrix Greens function $G^{-1}({\bf q}) = H_n({\bf q}) + H_h({\bf
q}) - E$.
Here $\phi({\bf q})$ denotes the Fourier transform
of the $4-$component wave function $\phi({\bf r})$.
Upon matrix inversion this can be rewritten as
${\bf M} \phi({\bf r}=0) = 0$,
with
\begin{equation}
{\bf M} = {\bf 1} + ua^2 \int {{d^2q} \over {(2\pi)^2}}
G({\bf q}) \tau^z \sigma^z   ,
\end{equation}
so that the eigenvalue condition is
the vanishing of the determinant:  $det({\bf M}) =0$.  Here
we are implicitly assuming that the integration is cut
off at high momentum by $q_c = 1/a$.

An explicit expression for the bound state
energy, $E_b$, can be readily obtained in the $u \rightarrow 0$ limit
by putting $E_b = \Delta_n + \Delta_h - \epsilon_b$
with {\it small} binding energy, $\epsilon_b$.  In this limit one need
only retain the contribution to the above integral
which is infrared divergent, which gives
\begin{equation}
{\bf M} = {\bf 1} + W(\sigma^z \tau^z - \sigma^x \tau^x )  ,
\end{equation}
with
\begin{equation}
W = -(u/8\pi \epsilon_0) \ln(\epsilon_b/\epsilon_0)  .
\end{equation}
Here we have defined an energy scale,
\begin{equation}
\epsilon_0 = (m_h v^2 + \Delta_n) /(2m_h \Delta_n a^2)  .
\end{equation}
Since $det({\bf M}) = 1-4W^2$, the eigenvalue condition
reduces to $W=1/2$, which gives the final result for
the bound state binding energy,
\begin{equation}
\epsilon_b = \epsilon_0 \exp(-4\pi\epsilon_0/u)   .
\end{equation}
Notice that the binding energy is exponentially small
in the interaction strength $u$, reflecting
the two-dimensional constant density of states for free massive holons
and nodons.  If one were to change the sign of
the interaction there is still a bound state (from $W=-1/2$)
with the {\it same} energy.
In either case, the bound state has the quantum
numbers of the electron with
$s_z =1/2$ and charge $\pm e$.  A spin down bound state
can also be readily found, corresponding
to the binding of a holon to a single nodon ``hole" in
the filled Fermi sea.

Between the threshold energy for generating the electron ($E_b$) and
the energy of the unbound nodon-holon continuum, the electron spectral
function should be governed by qualitatively the same physics as in
the AF case, except that the total spectral weight of the
corresponding feature will be reduced by matrix element factors
arising from, e.g. the possibly large spatial extent of the
nodon-holon bound state wavefunction.  Upon reaching the nodon-holon
continuum, we expect a much enhanced spectral weight but no sharp
feature at the continuum, as it is already lying in the continuum
formed by the electron plus spin-wave excitations, and the pair
excitation can thereby easily decay.

The $A({\bf k},\omega)$ in the AF and AF* phases are thus not
qualitatively different, and cannot strictly speaking be used to
distinguish the phases.  Quantitatively, however, we expect the AF*
spectral function to exhibit a very small ``quasiparticle'' peak, with
minimal separation from a nodon-spinon continuum carrying most of
the spectral weight.  If we assume that the holon gap
greatly exceeds the gap for the nodons,
then both features are expected to disperse in approximately d-wave
fashion, though the cusp for angles near $\pm 45^\circ$ should be
rounded by the nodon gap (Eq.~\ref{eq:nodongap}).  At non-zero
temperatures, thermally excited particles will scatter the injected
electron and lead to broadening of even the threshold peak.  In the
AF* case, where this feature is expected to lie close to the
nodon-holon continuum and have small weight, such thermal broadening
could well remove the quasiparticle peak completely at experimental
temperatures.

For a system at half-filling
but without particle-hole symmetry,
the Neel ordering
wavevector is not commensurate with the spacing between
antipodal nodes.  If this incommensurability
is sufficiently large, it is possible for the nodons
to remain gapless even in the presence of long-range
antiferromagnetic order.  In this unusual state,
which we denote as AF/NL, gapless
spin $1/2$ nodons co-exist with
the spin $1$ magnons.  Since the density of states
for the gapless nodons vanishes linearly with energy,
a weak interaction with the massive holons
is {\it not} expected to result in a holon-nodon
bound state.  Angle resolved
photo-emission in the AF/NL phase
will thus have a number of notable features.
Specifically, since the electron will decay
into the nodon-holon continuum, one does not
expect {\it any} sharp features in
the momentum resolved spectral function.
The lowest energy spectral
weight is expected
at the nodes, with a threshold
energy which disperses linearly away from the nodes
as in the d-wave superconductor.
The spectral weight should
rise smoothly above threshold due to the nodon-holon continuum -
with no delta function peaks.
This behavior is in fact reminiscent of
that observed in the undoped Ca compound by Shen et. al.,\cite{Shen}\
and is in marked contrast to the delta-function
spectral features expected in a conventional
antiferromagnetic.

\section{Discussion}

A small number of examples of condensed matter systems
are generally agreed to exhibit exotic quantum numbers, i.e. particles
which seem to require ``splitting'' the electron.  Both charge
fractionalization\cite{onedim} and spin-charge separation are generic in one
dimension\cite{Lutt}.  In the two-dimensional quantum Hall effect,
fractionally
charged particles have been known to exist for some
time,\cite{Laughlin}\ and recently
have been observed in dramatic shot-noise experiments.\cite{shotnoise}\
In both these
examples, fractional charge is connected to topological excitations:
solitons or domain walls in one dimension and vortices in two
dimensions.

A third example, less widely appreciated, is a superconductor in {\sl
  any} dimension.\cite{Kivelson90}\  For the superconductor the
mechanism is different:
{\sl Pairing} of electrons into singlets creates a gapless collective
(second) sound mode that carries the charge.  The sound mode can
adjust almost instantaneously to a quasiparticle, effectively
neutralizing it, leaving only a bare spin-$1/2$.  On the face of it
this species of spin-charge separation appears considerably different
from the other topological varieties.
\begin{figure}[htb]
  \epsfxsize=3.5in\epsfbox{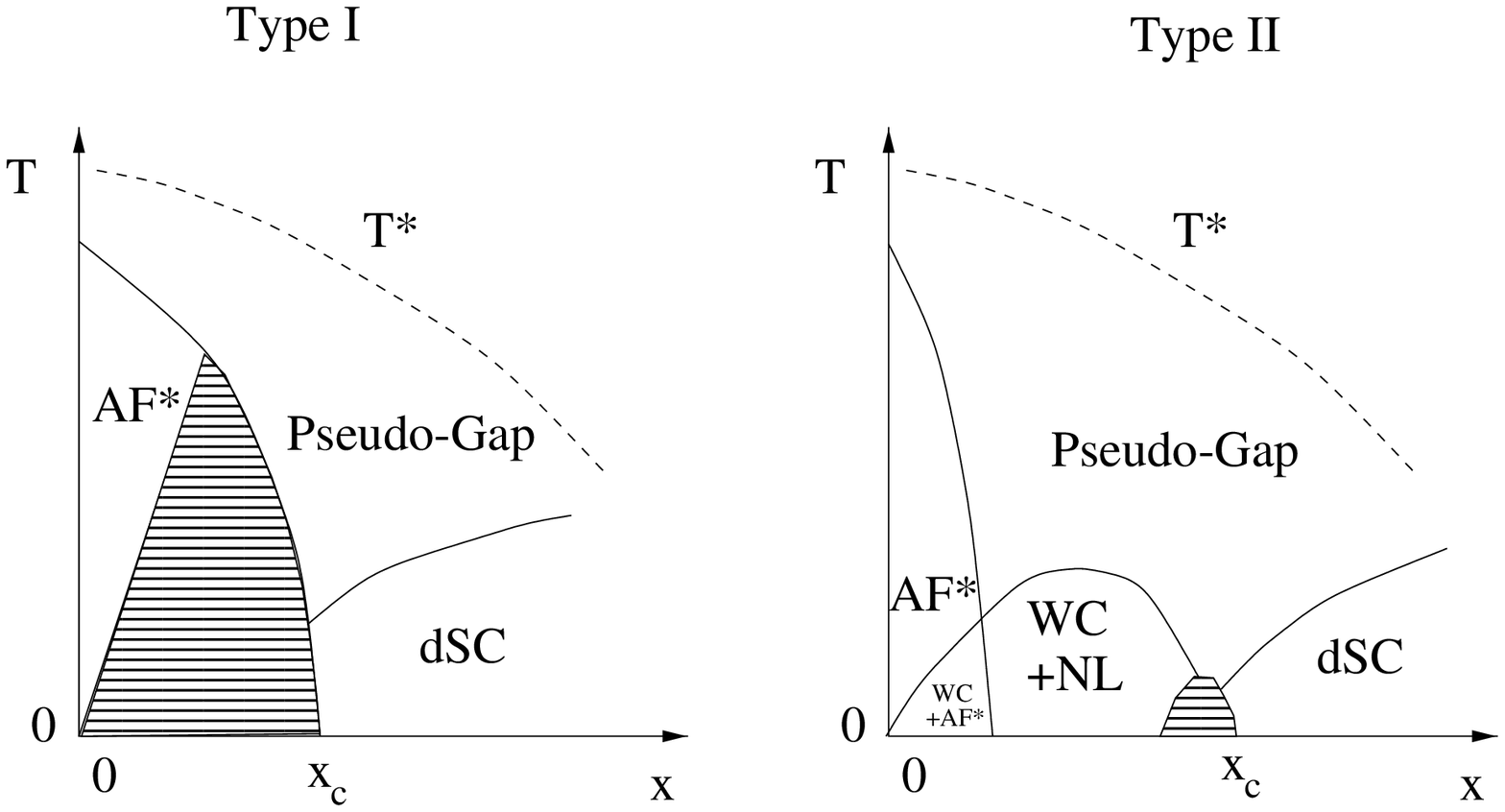} {Fig.~2:
    Phenomenological phase diagrams in the nodal liquid
    Ginzburg-Landau theory of Ref.~\onlinecite{BFN}.  Here $x$ is the
    hole doping and $T$ is temperature.  As discussed in the text,
    $hc/e$ vortex condensation leads to the unconventional AF*
    antiferromagnetic Mott insulator.  Note that, depending upon the
    magnitude of particle/hole asymmetry, nodons may remain gapless in
    an AF*/NL state at half-filling -- space constraints prevent us
    from indicating this on the figure.  Depending upon microscopic
    parameters, two principle phase diagrams occur opon doping.  In
    the type I scenario, added charge segregates into locally
    superconducting regions, which coalesce at some critical doping
    $x_c$.  In the type II scenario, added charges order into a Wigner
    Crystal (WC) with charge $e$ per period, presumably
    with some associated spin ordering.  After a small amount of
    doping the antiferromagnetic order is suppressed and the nodons
    are liberated into a nodal Liquid (NL) coexisting with the WC.  At
    $x_c$ this WC melts into the d-wave Superconductor (dSC) phase.
    See Ref.~\onlinecite{BFN}\ for details.  Some modifications are
    necessary if the effects of impurities are included, some of which
    are discussed in Ref.~\onlinecite{SFBN98}.}
  \label{pdfig}
\end{figure}
In this paper we have exploited a dual formulation to show that indeed
isolated charges (``holons'') derived from the superconductor can be
understood
as topological excitations in a vortex condensate. Further, we have described how spin-charge
separation can occur in an insulating state which results from the
quantum disordering of a superconductor.  Of course, propinquity to
the superconducting state doesn't guarantee the inheritance of
spin-charge separation: it only occurs when flux $hc/e$ vortices
condense.  The most interesting example of this phenomenon -- from the
point of view of high-$T_c$ phenomenology -- is the nodal liquid,
which we have discussed from this standpoint.  The condensation of
$hc/2e$ vortices, on the other hand, leads to the confinement of spin
and charge. The band insulator and the CDW, for example, can be
understood in this way.

One striking consequence of the distinction between spin-charge
separated and confined systems is that there are two distinct
antiferromagnetic states: one, AF, which is the ordinary
antiferromagnet and another, AF*, which is spin-charge separated.  The
latter results from ordering the nodons in a nodal liquid, so it has
neutral, spin $1/2$ excitations.  However, the distinction between the
AF and AF* phases is experimentally rather elusive. The natural place
to look is the electron spectral function, which can be probed through
angle-resolved photo-emission experiments. Under some conditions (see
Sec.~IVB), the unconventional antiferromagnet should exhibit {\sl
  only} a nodon-holon continuum instead of a quasiparticle pole.
Unfortunately, the existence of
nodon-holon bound states makes the distinction between the AF and AF*
phases rather subtle.  On the other hand, according to the paradigm
presented here, spin-charge separation is a consequence of $hc/e$
vortex condensation. Thus, spin-charge separation could be indirectly
evidenced by the observation of $hc/e$ vortices near the quantum
critical point at which superconductivity is destroyed.

The continuity of spin-charge separation that is embodied in the nodal
liquid and its offspring AF* state makes possible a simple
phenomenological description of the evolution from the insulator to
the superconductor, as espoused in Ref.~\onlinecite{BFN}.  For
example, the simplest Ginzburg-Landau formulation predicts the phase
diagrams in Fig.~2.  If, on
the other hand, spin-charge separation is {\sl absent} in the
undoped insulator, there must be a confinement transition between $x=0$
and the
superconductor.  While we have argued that such a
transition is driven by $hc/2e$ vortices, its nature and the phases
which it connects are highly non-trivial.  The relative simplicity and
elegance of the nodal liquid scenario thus argues in favor of its
relevance
to the cuprates.  Despite numerous and interesting differences among
different compounds, the phase diagrams of high-temperature
superconductors enjoy a remarkable degree of universality.  A number
of theoretical works have attempted to understand the commonalities
and variations amongst the topology of these phase diagrams
phenomenologically.\cite{so5}\ At low temperatures, however, we
believe classical phenomenology based only on conventional order
parameters misses the important physics of gapless quasiparticles and
spin-charge separation that are key in the vicinity of d-wave
superconductor-insulator
transitions.  {\sl Any} viable theory of the cuprates must at least
address the issue of how spin and charge either remain separated or
become confined on approaching the insulator.

%$+1$ dimensions, it is possible to create with finite energy a
%neutral, spin $1/2$ excitation at $x$ and a spin-less, charge $e$
%excitation at $x'$ even in the limit $|x-x'|\rightarrow\infty$.  These
%excitations only interact weakly, so correlation functions are
%convolutions of spin and charge pieces.  There is an additional effect
%in $1+1$-dimensions: all charges and spins move at the collective mode
%velocities, $v_c$ and $v_s$.  As a result, the convolution of charge
%and spin correlation functions is non-trivial, reflecting the break-up
%of electrons into excitations moving at these two velocities.

%When we contemplate spin-charge separation
%in higher dimensions, we see that the latter
%effect is fairly trivial but the former is not.
%If the electron somehow figured out a way to
%break up into spin- and charge-carrying excitations,
%which interact weakly with each other,
%these excitations would generically have different
%velocities.\footnote{Except, perhaps, at a quantum
%critical point at which these velocities might
%scale towards each other.}

In addition to the search for a `smoking gun'
experiment for spin-charge separation, there
are a number of other interesting questions
raised by this work. How do we implement
the interaction between nodons and $hc/2e$ vortices
in an $SU(2)$-invariant
way? How do these formulations of spin-charge separation
apply to $3D$ systems?

\acknowledgements

We are grateful to Eugene Demler,
Eduardo Fradkin, Steve Girvin, Courtney Lannert,
Subir Sachdev, Doug Scalapino and
Anirvan Sengupta for clarifying discussions.  We would 
particularly like to thank Bert Halperin for his insight
on the importance of gapped spin-charge separated excitations
as a means to  distinguish quantum phases.
This work has
been supported by the National Science Foundation under grants No.
PHY94-07194, DMR94-00142 and DMR95-28578.

\appendix

\section*{Dirac Lagrangian}

%\section{Bosonization Conventions}

%In our discussion of 1D models, we used the following
%bosonization conventions. The electron operator
%is decomposed into right- and left-moving components:
%\begin{equation}
%{c_\alpha}(x) = {\psi_{\alpha R}}\,{e^{i{k_F}x}} +
%{\psi_{\alpha L}}\,{e^{-i{k_F}x}}
%\end{equation}
%where $\alpha=\uparrow,\downarrow$.
%These have bosonic representation:
%\begin{equation}
%{\psi_{\alpha R,L}} = {e^{i\left({\phi_\alpha}
%\pm {\theta_\alpha}/2\right)}}
%\end{equation}
%where
%\begin{equation}
%[{\partial_x}{\phi_\alpha}(x),{\theta_\alpha}(x')] =
%-2\pi i \delta\left(x-x'\right)
%\label{commrel}
%\end{equation}
%We can introduce charge and spin bosons:
%\begin{eqnarray}
%{\phi_\rho} &=& {\phi_\uparrow} + {\phi_\downarrow}\cr
%{\phi_\sigma} &=& {\phi_\uparrow} - {\phi_\downarrow}\cr
%{\theta_\rho} &=& \left({\theta_\uparrow} +
%{\theta_\downarrow}\right)/2\cr
%{\theta_\sigma} &=& \left({\theta_\uparrow} -
%{\theta_\downarrow}\right)/2
%\end{eqnarray}
%which have the same commutation relations (\ref{commrel})
%with $\alpha=\rho,\sigma$.

\label{app:formalism}
Here we review the effective Lagrangian for low-energy d-wave
quasiparticles, following the notation of Ref.~\onlinecite{BFN}.  It
is most directly written in terms of the appropriate Nambu-Gorkov-like
spinor, $\Psi_{ia\alpha}$, with
\begin{eqnarray}
  \Psi_{i1\alpha}({\bf k}) & = & c_{{\bf K}_i+{\bf
      k},\alpha}^{\vphantom\dagger}, \\
  \Psi_{i2\alpha}({\bf k}) & = & i\sigma^y_{\alpha\beta} c_{-({\bf
K}_i+{\bf
      k}),\beta}^{\dagger},
  \label{Psidef}
\end{eqnarray}
where ${\bf K}_1$ and ${\bf K}_2$ are the momenta of the d-wave nodes
along the Fermi surface.  We use index-free notation in
which Pauli matrices $\vec{\mu},\vec{\tau},\vec{\sigma}$ act in the
node, particle-hole, and spin ($ia\alpha$) subspaces, respectively;
furthermore, if a single index is given explicitly, it is always the
node index.  In a particle--hole symmetric model at
half-filling, ${\bf K}_{1/2} = (\pm \pi/2,\pi/2)$ in the usual $(a,b)$
crystalline coordinate system (i.e. axes along the Cu-O bonds).  The
separation in Eq.~\ref{Psidef}\ is well-defined provided the momentum
is restricted to points near the nodes, i.e. $|{\bf k}|<\Lambda$,
where $\Lambda$ is a cut-off.

As in the s-wave case, we must allow for space-time dependence of the
superconducting phase $\varphi$.  For a d$_{x^2-y^2}$ superconductor,
one has
\begin{equation}
v\langle c_{i\uparrow}(t) c_{j\downarrow}(t)\rangle^\prime
= \Delta_d({\bf x}_i - {\bf x}_j)
\exp [i\varphi(\overline{\bf x},t)],
\end{equation}
where the prime on the angular brackets indicates an average (path
integral) over high-energy electronic states away from the nodes, and
$\overline{\bf x} = ({\bf x}_i + {\bf x}_j)/2$.  The amplitude
function $\Delta_d({\bf x})$ is the Fourier transform of the usual
momentum-space gap function, $\Delta_{\bf k} \sim f(|k|) [\cos^2 k_a -
\cos^2
k_b]$, and decays on the scale of $\xi$.     It is usually more
convenient for us to
work in rotated coordinates $x=(x_a+x_b)/\sqrt{2}$,
$y=(x_b-x_a)/\sqrt{2}$.  The appropriate effective quasiparticle
Lagrangian density was derived in Ref.~\onlinecite{BFN}:
\begin{eqnarray}
  {\cal L}_\Psi &=& \sum_{s=\pm} \Psi_1^\dagger \left( i \partial_t +
    iv_F \tau^z \partial_x + iv_{\scriptscriptstyle\Delta}\tau^s
    e^{is\varphi/2} \partial_y e^{is\varphi/2}
\right)\Psi_1^{\vphantom\dagger}\cr & &\, +
  (1\leftrightarrow 2;x\leftrightarrow y).
  \label{eq:qp}
\end{eqnarray}
Eq.~\ref{eq:qp}\ is derived on the assumption that the phase $\varphi$
is slowly varying on the scale of the coherence length,
i.e. $|\xi\partial_x\varphi|,|\xi\partial_y\varphi|,|\xi/v
\partial_\tau\varphi|\ll 2\pi$.  However, we expect on grounds of
universality that
Eq.~\ref{eq:qp}\ and its consequences provide a correct
low-energy description of the d-wave SC and its quantum-disordered
descendents more generally.\cite{foot3}\

The analysis of the interactions of vortices with quasiparticles is
based on the important change of variables
\begin{equation}
  \psi = \exp( -i\varphi\tau^z/2 ) \Psi.
  \label{eq:singular}
\end{equation}
Inserting Eq.~\ref{eq:singular}\ into
Eq.~\ref{eq:qp}, one finds ${\cal L}_\Psi = {\cal L}_\psi + {\cal
  L}_{\rm int}$, with
\begin{eqnarray}
  {\cal L}_{\psi} & = & \psi_1^\dagger [ i \partial_t + v_F \tau^z
  i\partial_x + v_{\scriptscriptstyle\Delta}
  \tau^x i\partial_y ] \psi_1 \nonumber \\
  & & + (1 \leftrightarrow 2,x \leftrightarrow y)  .
  \label{Lpsi}
\end{eqnarray}
The nodon field $\psi$ interacts with the phase of the order-parameter
as in Eq.~\ref{eq:Lint}.
Here the electrical 3-current $J_\mu$ is given by
\begin{equation}
  J_0 = {1 \over 2} \psi_j^\dagger \tau^z \psi_j^{\vphantom\dagger} ,
  \label{eq:J0}
\end{equation}
\begin{equation}
  J_j = -i{v_F \over 2} \psi_j^\dagger \psi_j^{\vphantom\dagger}  .
  \label{eq:Jj}
\end{equation}
Compared to the statistical gauge interaction with $\pm hc/2e$-flux
vortices, Eq.~\ref{eq:Lint}\ represents a much weaker {\sl two-fluid}
interaction between the quasiparticle or nodon current $J_\mu$ and the
superfluid current $\partial_\mu\varphi$.   A continuum duality
transformation appropriate for such a coupling was described in detail
in Ref.~\onlinecite{BFN}, and on the lattice in
Ref.~\onlinecite{Fisher98}.   Noting that the (Euclidian) 
nodon current $J^\mu$
couples to the
superfluid current $\partial_\mu \varphi$ in a manner directly
generalizing the Berry's phase coupling $in_0\dot\varphi$, the dual
(Euclidian) lattice action can be determined simply by replacing
$in_0\delta_{\mu 0} \rightarrow
in_0\delta_{\mu 0} + iJ_\mu$, i.e.
\begin{equation}
  {\cal S}_a \rightarrow \tilde{\cal S}_a = {u_2 \over 2} \sum_{j\mu}
(\epsilon_{\mu\nu\lambda}
  \Delta_\nu a_j^\lambda - N_0
  \delta_{\mu 0} - J_\mu)^2 .
  \label{eq:Satilde}
\end{equation}

The incorporation of antiferromagnetism was also described in
Ref.~\onlinecite{BFN}.  A low-energy effective Lagrangian describing
the magnon mode and its coupling to the nodons is
\begin{eqnarray}
  \label{Lneel}
  {\cal L} = {1 \over 2}K_{\mu} |\partial_{\mu}{\bf N}|^{2} -
  V_{N}(|N|) + g {\bf N}\cdot {\bf S}_{\vec{\pi}},
\end{eqnarray}
where $K_{0} = K$, $K_1 = K_{2} = -v_{s}^{2} K$, with $v_{s}$ the
spin-wave velocity in the AF.  Here 
\begin{equation}
{\bf S}_{\vec{\pi}} = {1 \over 
2}\left[\psi_j^{\dagger}\tau^{y}\bbox{\sigma}\sigma^{y}\psi_j^{\dagger} + 
{\rm h.c.}\right],
\end{equation}
is the spin operator at momentum $\vec{\pi}$.  Near any phase
transitions, and for most
phenomenological purposes, it is sufficient to take a simple form for
the potential: $V_N(|N|) = r_{N}|N|^{2} + u_{N}|N|^{4}$.  The
parameter $r_{N}$ controls the presence or absence of AF order.  In
mean-field theory, and neglecting for the moment the nodon coupling
$g$, the ground state passes from long-range to short-range AF order
as $r_{N}$ is tuned from negative to positive.  We include only the
most relevant coupling of the N\'eel
field to the nodons allowed by symmetry,
\begin{equation}
  {\cal L}_{nodon} = {\cal L}_\psi + g{\bf N}_0 \cdot {\bf S}_{\vec{\pi}} ,
\end{equation}
with ${\bf N}_0 = \langle {\bf N} \rangle$. 

A compelling feature of the above description is the resulting
low-lying spectrum in the antiferromagnet.  The model can be readliy
diagonalized with an appropriate Bogoliubov transformation, giving
the energy eigenvalues,in Eq.~\ref{eq:nodongap}.  In all
nodon sectors there is a non-zero {\it gap},
equal to $g N_0$.  The nodons having been lifted to finite energy, the
only remaining gapless excitations in the AF* phase are the spin waves (slow
rotations of ${\bf N}$) dictated by Goldstone's theorem.

\end{multicols}


\begin{references}
\bibitem{foot1} There are actually two loopholes.  The first is if the
  effective action contains a Chern-Simons term for $a_\mu$. This can
  only happen in a system without time-reversal symmetry, and the
  result is the quantum Hall effect. The other possibility occurs if
  the effective action for $a_\mu$ contains a non-local term, such as
  ${\left({\partial_\mu}{a_\nu}\right)^2}/{\partial^2}$.  This can
  only happen if there are gapless degrees of freedom coupled to
  $j_\mu$, as in a metal. It is hard to see how such a system can be
  insulating, but we cannot rule it out.
\bibitem{foot2} We cannot rule out the possibility that
  translational symmetry is broken through the development of
  ``staggered flux phase'' ordering, $\langle {c^\dagger_{\alpha
      k+Q}}\, {c_{\alpha k}^{\vphantom\dagger}} \rangle = \cos {k_x} - \cos {k_y}$, in
  which case no gap is opened.
\bibitem{foot3} It is of course possible that different physics
  neglected here could obtain in the vortex cores and lead to distinct
  quantum-disordered insulating states with additional degrees of
  freedom.  There appears, however, no compelling argument why this
  {\sl must} be the case, particularly given the extremely short
  coherence lengths of the high-T$_c$ materials.  We therefore leave
  the incorporation of ``core states'', should they exist, as an
  interesting open problem, focusing here on the more fundamental
  physics in their absence.
\end{references}
\end{document}